# Application of Artificial Intelligence and Machine Learning in Libraries: A Systematic Review

Rajesh Kumar Das
*Noakhali Science and Technology University*, rajesh@nstu.edu.bd

Mohammad Sharif Ul Islam
*University of Dhaka*, sharif.islm@du.ac.bd

# Application of Artificial Intelligence and Machine Learning in Libraries: A Systematic Review

## Introduction

Over the past few decades, Artificial intelligence (AI) and Machine Learning (ML) have become major catalysts of reshaping our world and the way we think, act and make decisions (Vysakh & Babu, 2020). Recently, many different attributes of machine learning and artificial intelligence have been adopted by many leading organizations such as Google, IBM, Amazon, Netflix, Expedia and so on to improve their products and services. Almost all major sectors like; health, education, weather, business, stock, agriculture, government and non-government agencies of different countries are also showing interest and using these technologies to simplify and neutralize workload, increase and speed up productivity, reduce human interaction and most importantly lead the digital-world in a smart and sophisticated way.

Like any other fields, libraries and information sectors are also incorporating many fascinating technologies into their armories. This is because of the presence of ever-increasing volumes of data, which is often referred to as big data; the requirement of real-time data processing and generating results; and the diverse needs of the library patrons always pushing the library and information sectors to the edge. On the other hand, the major advancement in computer processing speed and capacity; popularity of using the networked environment for data processing etc., are some of the major potential combinations that create the possibility to mining real-time data and deliver information outputs accordingly (Johnson, et. al., 2015). For instance, Application of AI and ML is helpful for better interaction among smart technologies to enhance the effectiveness of different libraries that ultimately shifts the traditional library services to intelligent library systems, by revolving users' needs and providing customizable and ubiquitous knowledge services. Subsequently, ML methods that provide computational solutions for automatically acquiring new knowledge become crucial for the development of a truly intelligent library (Esposito et al. 1998), which is equally effective irrespective of time, place, or location (Zhiyong, 2019).

Thus, it can be argued that the introduction of AI & ML has created a new horizon in revolutionizing both technical and user services in libraries. Self-learning and self-doing ability of AI and ML can help libraries for better interaction among machine-automated intelligent technologies for the effectiveness and co-creation of all library services. However, to cope up with the transformed scenario, librarians have to change their roles and promote the transformation of library operations and services assisted with machine learning and artificially intelligent technologies. With the development effort of the aforementioned smart technologies, a wide range of research has been conducted for understanding the phenomena and creating innovation in this field. So, to trace the development and intellectual structure of a knowledge domain, it is necessary to know about the present research focus and thus visualizing the future of a particular domain. Therefore, this study seeks to understand the current state of the art of the AI & ML applications into libraries and to predict where future research will lead.

## Background

### *Artificial Intelligence*

Artificial intelligence (AI) is the ability enabled by a digital computer or computer-controlled machine or software replicating intellectual characteristics like intelligent organisms (human) in their functionality. Major AI scholars and textbooks define this field as the design and development of "a fully conscious, intelligent, computer-based entity" (Raynor and Shoup, 1999) that has intrinsic advantages over human in perceiving the environment and maximizing the success of complex tasks (Russell & Norvig, 2003). John McCarthy, who coined the term in 1955, defines AI as "the science and engineering of making intelligent machines" (McCarthy et al., 2007:2). The central goals of AI are to reason, discover, generalize, manipulate objects and natural language processing etc. (Nilsson 1998; Poole, Mackworth & Goebel 1998; Russell & Norvig 2003; Luger & Stubblefield 2004; Copeland, 2015). AI has been the subject of great enthusiasm in recent years in varied disciplines like computer science, psychology, mathematics, information science, linguistics, and other specialized domains. In the case of LIS, the most extended AI presence has been identified as the appearance of the expert system. The application of expert systems not only helps the library professionals in performing the basic library operations (Guliciuc et al., 2017) but also helps in decision-making process and improvement in productivity.

AI has the ability to think and act like a human without any human interference, it can help in the evolution of an intelligent library with latent intelligent roles to perform without the intrusion of human support (Massis, 2018). Self-learning ability of AI can prove very important to libraries in terms of user handling, networking and communication (Huang and Rust, 2018). AI technologies also could be used to provide innovative real-time virtual reference services through mobile and social networking environments, by combining the existing library resources and third party contents. Additionally, some other promising areas of AI in libraries include natural language processing, indexing systems, and application of robotics in library activities.

### *Machine Learning*

The term 'Machine Learning' was first coined by Arthur Samuel in 1959 after Alan Turing's lecture at the London Mathematical Society back in February 20, 1947 where he said that we want such a machine with the ability to learn from experiences (Samuel, 1959). Thus, Samuel defined Machine learning or Ml as a "field of study that gives computers the ability to learn without being explicitly programmed" (Samuel, 1959, pp. 210–229). Ex Libris Whitepaper (2020) defines Machine Learning as "…when machines create their own classifications by learning from examples, dramatically accelerating statistical pattern recognition. ML studies and develops algorithms by predicting on data through uncovering complex patterns for making intelligent decisions (Foster and Kohavi, 1998). Thus, in layman's terms, machine learning involves using computers (i.e., *machines*) to identify patterns within large amounts of data (Ayyadevara, 2018; de Mello & Ponti, 2018). As the name suggests, machine learning implies that, over time as experience accrues (i.e., a larger sample of data analyzed), the computer will become better at performing analytical tasks (Bellam, 2018; de Mello & Ponti, 2018). However, what makes machine learning possible are *learning algorithms*, which facilitate one of the two main learning model approaches—*supervised* and *unsupervised* or *non-supervised* learning (Ayyadevara, 2018; Fernandes de Mello & Antonelli Ponti, 2018). These algorithms can be described simply as; step-by-step instructions that allow a computer to solve a particular type of learning problem (Bellam, 2018).

Machine learning has now been considered as a game-changer through reaching out solutions to complicated real-world problems in a scalable way useful with a wide range of computing tasks. Machine learning is sometimes fluxed with data mining where machine learning concentrates on prediction on the basis of known properties that are learned from the training data; and data mining on the other hand, focuses on the discovery of unknown properties in the data. The question is how Machine learning technologie are applicable to libraries? Short answer to this question is, the opportunities and possibilities are limitless. For instance, ML techniques could be useful for resource discovery. The web crawlers and other data harvesting tools can be applied for automation and advanced classifications of information resources to redefine the resource accessibility of library users (Mitchell, 2006). Data mining techniques can also be used to harvest data from both homogenous and heterogeneous systems for understanding the library usages patterns (Walker & Jiang, 2019).

## Research Question

Two research questions were defined and applied to the selected studies. The research questions are as follows-

- *RQ 1:* What are the main areas of library where AI and ML have been applied?
- *RQ 2:* Which AI and ML techniques are applied in libraries?

## Methodology

This study has been adopted systematic review process proposed by Kitchenham et al. (2009). Systematic review is a well-defined and structured process that appraises and synthesizes best available studies on a specific topic. It provides evidence based answers to the specific questions using explicit, accountable, and rigorous research methods (Gough, Oliver & Thomas, 2012; Dickson, Cherry & Boland, 2014; Petticrew & Roberts, 2006). The main aim of the systematic review is to provide a comprehensive, exhaustive and complete summary of current evidence drawn from the literature, in an unbiased and reproducible way. It helps to provide evidence for practice and policy-making and identify gaps in the research (Siddaway et al., 2019).

This study has been motivated by several factors. Artificial Intelligence (AI) and Machine learning (ML) approaches have been utilized across a range of applications in libraries. The increasing attention of AI and ML towards library sector has been guiding the growth of research practices in this domain. This trend has buttressed the need for a systematic review of literature to summarise the ML and AI methods used for the varied application areas of libraries. In our view, the rapid development of this new trend requires a comprehensive review that will trace the evolution of this field, and as well as will explain the applications of AI and ML in various sub-domains of libraries.

### *Data Source and Search Strategy*

A comprehensive literature search was carried out in October 2020 in Web of Science, Scopus, LISA and LISTA databases. These databases were used as they are commonly known and widely used by the scientific community for their scholarly communications. Besides, LISA, LISTA have been trusted sources of LIS research since the very past. Individual database searching was also done across various databases to avoid search biases. The search was performed for all articles published to date and sorted by most recent to least recent. The search was restricted to peer-reviewed journals, conference papers, and proceedings.

Table 1

Database and search strategy

| **Database** | **Limiters** | **Search string** | **Source** | **Search date** | **Documents** |
|---|---|---|---|---|---|
| WoS | Title, abstract, keywords | “machine learning” OR "artificial intelligence" AND “librar*” | Conference Papers & Proceedings, Scholarly Journals | 8/10/2020 | 9 |
| Scopus | Title, abstract, keywords | | | 8/10/2020 | 26 |
| LISTA | Title, abstract, keywords | | | 13/10/2020 | 21 |
| LISA | Title, abstract, keywords | | | 13/10/2020 | 285 |
| Individual Database Searching | Title, abstract, keywords | | | 10/10/2020 | 38 |

***Inclusion and Exclusion Criteria***

Inclusion criteria were: (1) articles must be written in English, (2) articles must be about ML or AI applications, (3) articles must belong to the sub-area of Library and Information Science (4) title, abstract and keywords of the articles must discuss libraries with the application of either machine learning or artificial intelligence. The exclusion criteria were: (1) non-digital publications, (2) Publications not available for full review, (3) articles not in English language, (4) There is no application of ML or AI in libraries.

***Study Selection***

Following the established strategy, an early screening of the articles based on title and abstract were performed by researchers. Afterward, the remaining articles were assessed by applying the aforementioned inclusion and exclusion criteria. Initially, 379 articles were retrieved (9 from Web of Science, 26 from Scopus, 21 from LISTA, 285 from LISA databases and 38 from individual database searching) with the established Boolean

Expression. After removing 61 duplicates (16.09%), the search was reduced to 286 articles. After careful screening of the remaining 286 papers considering the inclusion and exclusion criteria, 32 articles were included in this systematic review. The selection procedure is summarized in Figure 1.

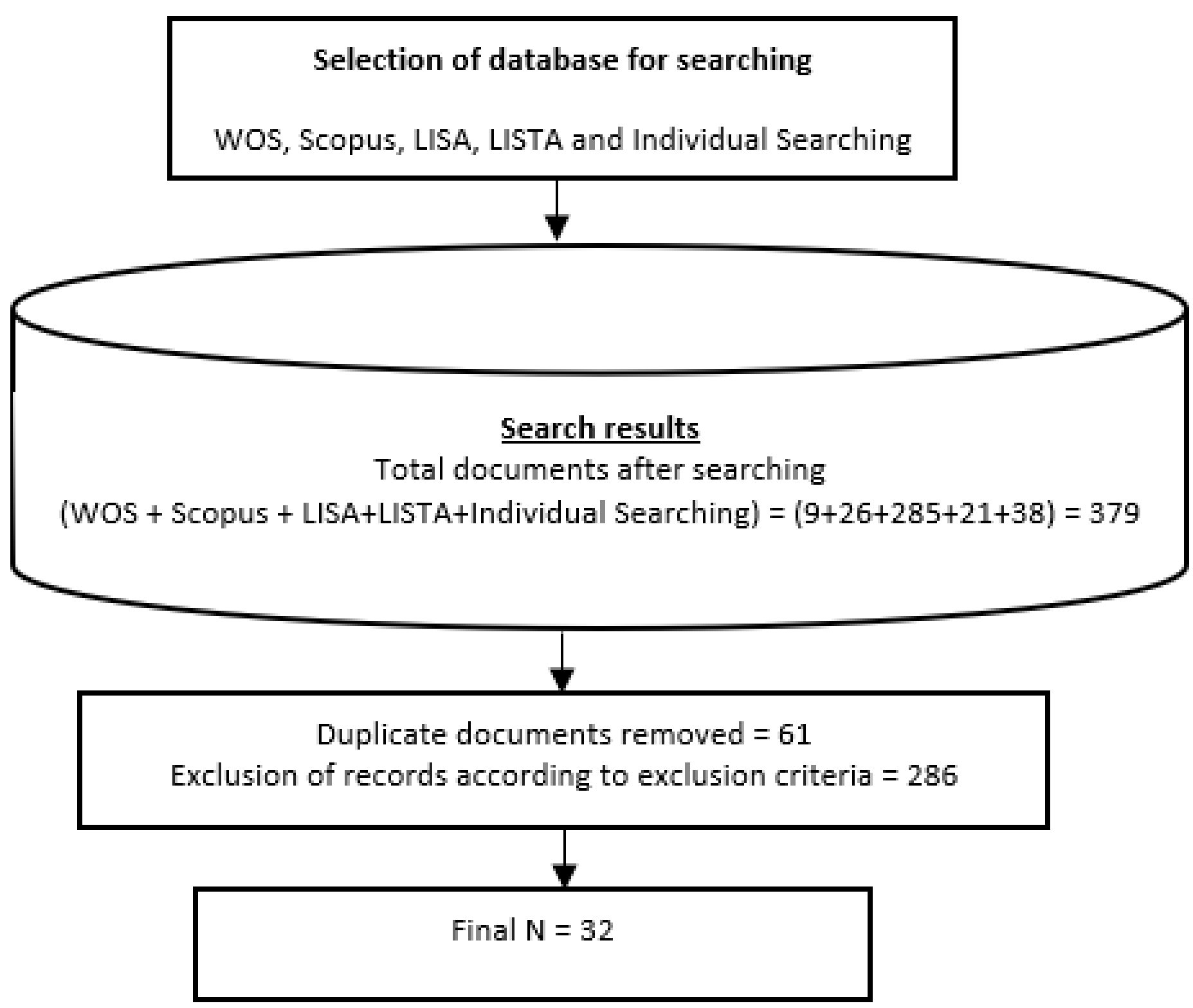

Figure 1

Search Process for the Selection of Studies

## Findings

### *AI and ML Techniques Applied in Libraries*

As mentioned previously, after careful consideration, a total number of 32 articles have been finally selected and included for this study. The papers went through a rigorous data extraction method done by the two researchers. The relevant theme related to the use of AI and machine learning in different aspects of library operations were identified and summarized in the table-2.

Table 2

Application of AI and ML techniques in libraries

| Theme | Task in libraries | Machine learning techniques | Artificial intelligence technique | References |
|---|---|---|---|---|
| Collection building and management | i. Metadata generation, resource discovery<br>ii. Book acquisition | i. Logistic Regression, kNearest Neighbor<br>ii. Adaptive boosting, Regression | | i. Mitchel, 2006<br>ii. Walker & Jiang, 2019; Omehia, 2020 |
| Circulation and user service | i. Book recommendation<br>ii. Reader ratings, bibliographic data | i. Association rule, Support Vector Machine (SVM)<br>ii. Recommender System | | i. Tsuji et al., 2014<br>ii. Xiao & Gao, 2020 |
| Processing in libraries | i. Document analysis, processing, classification, document understanding, text recognition<br>ii. Indexing, cataloguing, subject analysis | i. Preprocessing | i. Natural language processing<br>ii. Text data mining, clustering | i. Esposito et al., 1998<br>ii. Short, 2019; Boman, 2019 |
| Reference service | i. Virtual reference service, readers' advisory librarians, and virtual story-tellers<br>ii. Translation of text | | i. Self-directed learning<br>ii. Natural language processing | i. Yao et al., 2015; Rubin et al., 2010; Cox et al., 2019a; Gul & Bano, 2019<br>ii. Yao et al., 2015 |

| Library administration | i. Library security; User identification; Book title recognition<br>ii. RFID management | i. Pattern recognition | ii. Multi-agent system (MAS) | i. Yang et al., 2017; Lei et al., 2018; Du et al., 2019<br>ii. Minami, 2008; |
|---|---|---|---|---|
| Library Customization and retrieval | i. Information searching, retrieval | i. Deep learning | | i. Yang et al., 2017;<br>Liu (2011); Kumar & Srinivas, 2004; Wójcik, 2020 |
| Research and scholarship | Theoretical paper | | | Massis (2018); Fox, 2016; Allen et al., 2002; |
| User study on the of ML&AI implemented in libraries | Case study, survey, interview etc. | | | Chen & Shen (2019); Zuccala, et al., 2007; Cox, et al.2019b; Fernandez, 2016; Rubin, et al., 2010; Nardi & O'Day, 1996; Frias-Martinez et. al., 2006 |
| Existing AI & ML technologies and solutions in Libraries | Review | | | Mack, 2002; Asemi, et al., 2020;<br>Xiao & Gao, 2020; Schreur, 2020; Arlitsch & Newell, 2017 |

Based on the findings (Table-2), Recommender systems (RSs) could be defined as software tools and techniques that generate meaningful recommendations to the users based on their likings of items and products. The recommender systems provide a personalised experience to the users through filtering, rating, preference or options that might interest them as output (Burke, 2002). RSs are becoming popular tools for evaluating and filtering large amounts of information too. In order to improve library services and provide personalised services to the users, different kinds of such recommendation algorithms have been widely used by the libraries (Xiao & Gao, 2020). For instance, Tsuji et al. (2014) used 2,293,642 loan records from 44,571 users of the University Library to recommend books. They used two machine learning techniques; (i) SVM and (ii) association rules to recommend books to both students (undergraduate and graduate) and faculties. SVM was used according to: (a) the confidence from an association rule, (2) likenesses between titles, (3) match/mismatches between categories, and (4) similarities between the outlines in the book databases. Xiao & Gao (2020) utilize Bayes estimator to offer generalized recommendations based on the item's popularity score. Hence, popular books have a greater probability of being chosen by other readers. Walker & Jiang (2019) employed logistic regression and AdaBoost to investigate the viability of machine learning as a platform for predictive modelling of demand-driven acquisition, (DDA) and purchasing of e-books. The logistic regression analysis was performed by general linear model (glm) R function and the Ada-Boost was powered by the adabag R package. Concerning the implementation of real-time virtual reference service by a newly developed smart talking robot called Xiaotu, Yao et al (2011) employed self-directed learning techniques. Short (2019) experimented with text mining, classification, clustering and topic-modelling focusing on 5 years-subject analysis of a novel to enhance the experience of cataloguing and subject analysis. Natural language processing (NLP), a branch of AI, was also employed in this study to understand the Chinese natural language talking function through natural language communication between respondent and robot. Another sophisticated AI technique, pattern recognition investigated by Yang et al. (2017). This study focused on book inventory management of a library and the retrieval system based on scene text reading. They proposed a design text recognition model utilizing rich supervision. Lei et al. (2018) proposed a convolutional-neural-network-based book label recognition approach for digging out the misplaced books. The authors emphasized on image processing techniques and convolutional neural networks (CNNs) to excerpt the label's characters attached to each book from the images of the bookshelves, and to train a classifier for recognizing characters respectively. In a recent study, Du, Lim & Tan (2019) emphasized on library security through passive RFID tags based on pattern recognition through collection and analysis of profiling distribution, reading activity, reader's trajectory, and thus identification of picked up and misplaced books.

### *Research Themes and Areas of Libraries*

By reviewing the use of AI and ML for libraries in the selected articles, a total number of ten themes emerged, these are; collection building and management, processing in libraries, circulation and user studies, reference service, library administration, library customization and retrieval, research and scholarship, service quality and innovation, intelligent agents for information search and retrieval, study on implementation and existing technologies and solution. Table 2 summarizes these applications and categorizes the references with respect to the literature features they have investigated.

#### *Collection Building and Management*

Through AI, it is possible to discover, search and analyse any vast library collections. It brings the skills and knowledge of library staff, scholars, and students together to design an intelligent information system that respects the sources, engages critical inquiry, fosters imagination, and supports human learning and knowledge creation (Massis, 2018; Cox et al., 2019). In the process, the Service innovations will double the result as it will cut the effort half for the intelligent library's development and will be the most important development direction of the intelligent library (Chen & Shen, 2020).

*Circulation and User Service*

In addition, the services provided by the intelligent library have a great impact on the frequency of using the resources of the library for readers (Chen & Shen, 2020). Significantly library use data would feed into learning analytics. The data that libraries hold on user activity and the way they use services will contribute more widely to their kind of learning journey in the future (Cox et al., 2019).

*Processing in Libraries*

Some of the major activities for any libraries include acquisition, processing, collection building and management. AI helps with potential intelligent roles like, data acquisition, data curation, and data quality control (Gul & Bano, 2019). Moreover, AI can be utilized in cataloguing, classification, acquisition of collections, indexing and management activities as well (Walker & Jiang, 2019; Wójcik, 2020; Omehia, 2020). On the other hand, Machine learning assists in libraries for metadata generation, resource discovery, and rich full-text identification and extraction (Mitchel, 2006; Cox et al., 2019; Limb, 2002). Machine learning is used to generate subject headings, for example, metadata directly describing the resource (a bibliographic record) and supplemental metadata about the author or subjects (Boman, 2019; Short, 2019). Different intelligent technologies are also used for improving better library operations and services through AI and ML technologies, such as intelligent shelves. Intelligent shelf is capable of reading books on the shelves, although the shelf itself cannot tell the specific location for a book. However, an intelligent browsing table in the Library, which is able to collect information about users' reading habits and this information, can be used in the arrangement of collections and shelves (Minami, 2005).

*Reference Service*

Another important aspect of library services focuses on reference services, and here also ML and AI can play major roles. An artificially intelligent conversational agent or chatbot might works as a virtual reference librarian which enhance face-to-face human interaction, and also act as library web site tour guides, automated virtual reference and readers' advisory librarians, and virtual story-tellers (Rubin et al., 2010; Yao et al., 2015, Cox et al., 2018; Gul & Bano, 2019). Some examples of virtual/dynamic reference services include; 'the Stanford Encyclopaedia of Philosophy project' in which an expert team offers dynamic reference services (Allen et. al., 2002), 'Virtual Reference Desk' in which reference agents deliver digital reference services (Lanks, 1999) and an agent-based system that supports access to information from the repositories in the web (Rizzo et al., 1998).

*Library Administration*

Besides, some autonomous agents develop their own specifications after watching the user's interests, while some other autonomous agents find out the dead links in the digital libraries and try to fix them or inform the administrator (Kumar & Srinivas, 2004).

*Library Customization and Retrieval*

Another major area of implementing AI and ML is information search and retrieval. Agent technology has been used to support the information search process in DLs, including strategic search support, proactive support for query formulation, intelligent assistance to information search and retrieval, and personalised services to users (Liu, 2011). An agent-based architecture supports high-level search activity in federated DLs combining the advantages of previous approaches (Wójcik, 2020). Depending on the interest of the user, intelligent software agents search the DLs and return the information which the user is looking for and also through the personal agents, users can customize their interfaces (Kumar & Srinivas, 2004). Therefore, librarians now tend to use artificially intelligent solutions for advancing the better, quicker and efficient process of acquisition, processing and analysis of information to meet information needs of users (Wójcik, 2020).

*Research and Scholarship*

According to the literature, AI could shape the library operations and services with the focus on the library's basic operations, administrations, research, scholarship, service innovation, usability, retrieval and so on. Deep learning, neural network algorithms, convolutional neural networks have also been proved as powerful tools for scholarship and research reported in the study (Massis, 2018; Fox, 2016; Allen et al., 2002).

*User Study on AI and ML Implemented in Libraries*

There have been several studies found in the review that investigated the behavior of users from and machine-enabled and intelligent library. Nardi and O'Day in their 1996 paper appears to be the first to introduce the concept of an intelligent reference librarian for user service through the application of artificial agent in libraries. As shown in the literature, user modelling in digital libraries is also being done through machine learning by following the process of data preprocessing, pattern discovery and validation (Frias-Martinez et. al., 2006, Zuccala, et al., 2007). User study on the application of conversational agent inside and outside libraries including academic, informative, assistive activities for virtual advisory service, reference service to the users (Fernandez, 2016; Rubin, et al., 2010). Service quality of intelligent library are also being measured from different perspective like technology acceptance, research data management etc. (Chen & Shen (2019; Cox, et al., 2019b).

*Existing AI & ML Technologies and Solutions in Libraries*

Several papers related to the AI & ML technologies and their solutions in Libraries were found important as to support a variety of library operations like reader rating, resource discovery, delivery and preservation through the use of recommender systems, linked data, expert systems, robots etc. (Mack, 2002; Arlitsch & Newell, 2017; Asemi, et al., 2020; Xiao & Gao, 2020; Schreur, 2020).

**Discussion and Conclusion**

The application of artificial intelligence and machine learning in libraries is an emerging trend that has captured the attention of relevant practitioners and academics. The aim of this systematic review has been to identify the application of artificial intelligence and machine learning in libraries, while assessing how these technologies could assist and support the library operations and services. Thirty-two papers were identified, analyzed and summarized on the application of AI and ML domain and techniques which are most often used. Although this review cannot claim to be exhaustive, it does highlight important implications and provide reasonable insights:

The current state of the AI and ML research that is relevant with the LIS domain mainly focuses on theoretical works. However, some researchers also emphasized on implementation projects or case studies. According to the literature, AI could shape the library operations and services with the focus on the library's basic operations, administrations, research, scholarship, service innovation, usability, retrieval and so on. For collection management in libraries, several Ml techniques like logistic regression, KNN, AdaBoost have been widely used for Metadata generation, resource discovery; and Book acquisition. Whereas for circulation (book recommendation, user rating, bibliographic data etc.) recommender system, SVM, association rule have been utilized. Library in-house activities like; cataloguing, classification, indexing, document analysis, text recognition etc., have been supported by both AI and ML technologies.

Some advanced AI and ML techniques like pattern recognition and MAS are also being used to ensure library security; user identification; book title recognition; RFID management, and other administration activities. Deep learning, neural network algorithms, convolutional neural networks have also been proved as powerful tools for scholarship, collections discovery, search and analysis.

Besides, an artificially intelligent conversational agent or chatbot works as a virtual reference librarian. It enhances face-to-face human interaction for library web site tour guides, automated virtual reference assistants, readers' advisory-librarians, and virtual story-tellers. In any case, as a systematic review study this research also has some limitations. Firstly, the study examines papers extracted based on specific keywords such as "machine learning" and "artificial intelligence" and "librar*". Articles which mentioned the application of AI and ML techniques in libraries without these keywords may have been omitted during the retrieval process. Secondly, findings are based on data collected only from academic journals and conference papers, so other materials which may contain more studies on issues might have been excluded. Thirdly, articles from the limited number of databases (2 multidisciplinary and 2 LIS), were included. However, although this limitation could mean that the review is not exhaustive, the authors believe that it is comprehensive by providing reasonable insights into the work being accomplished in the LIS research domain. Despite the limitation, this study could help in the development of new ideas and models or tools to support and enhance the existing service ecologies of libraries. This study will provide a panoramic view of AI and ML in libraries for researchers, practitioners and educators for furthering the more technology-oriented approaches, and anticipating future innovation pathways.

**Author Contributions**

**Rajesh Kumar Das:** Conceptualization, Methodology, Writing, Analysis

**Md. Sharif Ul Islam:** Methodology, Validation, Editing

**References**

**Asterisk * indicates the articles included for the systematic review**